\documentclass[prl,twocolumn]{revtex4}
\usepackage{epsfig}
\usepackage{float}
\usepackage{times}

\def\bK{\mathchoice{\mbox{\boldmath $\displaystyle K$}}%
                   {\mbox{\boldmath $\textstyle K$}}%
                   {\mbox{\boldmath $\scriptstyle K$}}%
                   {\mbox{\boldmath $\scriptscriptstyle K$}}}

\def\bE{\mathchoice{\mbox{\boldmath $\displaystyle E$}}%
           {\mbox{\boldmath $\textstyle E$}}%
           {\mbox{\boldmath $\scriptstyle E$}}%
           {\mbox{\boldmath $\scriptscriptstyle E$}}}

\def\imu{{\rm i}}
\def\eun{{\rm e}}
\def\RE{\mathop{\rm Re}\nolimits}

\def\am{\mathop{\rm am}\nolimits}
\def\abbr#1{{\rm #1}}

\newcommand{\be}{\begin{equation}}
\newcommand{\ee}{\end{equation}}
\newcommand{\bea}{\begin{eqnarray}}
\newcommand{\eea}{\end{eqnarray}}
\newcommand{\rd}{{\rm d}}
\newcommand{\p}{\partial}

\newcommand{\thalf}{{\textstyle\frac{1}{2}}}

\begin{document}

\title{Metal--Insulator Transition Tuned by External Gates in Hall Systems with Constrictions}

\author{Emiliano Papa$^1$ and Tilo Stroh$^2$}
\affiliation{$^1$Department of Physics, University of Virginia, Charlottesville, 
VA 22904-4714\\
$^2$Department of Physics, University of Siegen, 57068 Siegen, Germany}

\date{\today}

\begin{abstract}
The nature of a metal--insulator transition tuned by 
external gates in quantum Hall (QH) systems with point constrictions, as reported 
in recent experiments \lbrack S.Roddaro {\it et al.}, Phys. Rev. Lett. {\bf 95}, 156804 (2005)\rbrack, is examined. 
We attribute 
this phenomenon to a splitting of the integer edge into conducting and insulating stripes, 
the latter wide enough to allow for the stability of the edge structure. 
Inter-channel impurity scattering and inter-channel Coulomb interactions do not destabilize 
this picture.

\end{abstract}

\maketitle

{\it Introduction}---A current  interest in one-dimensional electron systems is driven
in part by the  advances in technology which  have enabled fabrication
of new materials and in part by a hope to observe the exotic effects predicted
theoretically.
In this context, the edge states of QH systems offer an
especially attractive terrain for testing the properties of one-dimensional fermion states.

 Even though remarkable theoretical progress has been achieved in
understanding properties of Luttinger liquids (LL) \cite{gogolin,voit},
 some open problems remain on the experimental front \cite{grayson,heiblum}. 
For instance, there are problems   concerning  the transport measurements
across a constriction in a two-dimensional electron gas (2DEG) conducted by Roddaro {\it et
al.}\ \cite{pellegrini}.
The theory \cite{kane,fendley} predicts
a change in the nature of linear
transport across a backscattering point,
depending on the value of the  LL parameter:
perfect transmission for attractive interactions across the sample
and perfect reflection for repulsive interactions.
For QH systems of bulk filling $\nu$, at finite voltage the theory predicts the
power-law left--right tunneling differential conductance
 $G\sim V^{2/\nu -2}$.
Contrary to these expectations, measurements in QH 
samples at integer filling \cite{pellegrini} reveal a metal--insulator transition across the point contact,
tuned by the voltage of the top metallic gates, $V_\abbr{g}$.
The system is metallic at low $\left|V_\abbr{g}\right|$ 
but becomes insulating at high $\left|V_\abbr{g}\right|$.
The bias-voltage evolution of constriction transmission in the latter case is similar
to the one corresponding to bulk filling $\nu=1/3$.
\looseness=-1

The low $\left|V_\abbr{g}\right|$ behavior was examined in \cite{papa} and it
 was found that interedge repulsions 
across the gates can lead to 
suppression of quasi-particle backscattering at the constriction.
In analogy, the insulating behavior  of \cite{pellegrini} at higher $\left|V_\abbr{g}\right|$ 
(backscattering enhancement) would have been understood if electron interactions reversed sign 
at high $\left|V_\abbr{g}\right|$, due in this case to the presence of image charges on the gates.
Allowing for different potentials $V_1,V_2$ on the two 2DEG subsystems, and studying the 
change of the position of
one of the edges upon variation of both of these voltages \cite{macd}, we find that electron 
interedge interactions remain repulsive at any values of $V_1$ and $V_2$.
\looseness=-1

We examine then the structure of the QH edges 
surrounding the gates in presence of the magnetic field.
As in the semi-infinite-gate case \cite{glazman,chklovskii}, a sequence of compressible (CS) and incompressible (IS) 
stripes arises, but the geometry of these systems and $V_{\rm g}$ in
\cite{pellegrini} are nevertheless such the ISs and the CSs have comparable widths. 
The presence of ISs strongly influences the tunneling exponents, leading eventually to
suppression of electron tunneling across the constriction.
\looseness=-1

\begin{figure}
\begin{center}
\epsfig{file=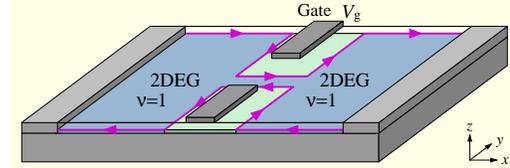,width=68mm} 
\end{center}
\vskip-\lastskip
\caption{Schematic illustration of a quantum Hall bar with a split-gate
constriction. The gates bring counter-propagating edges in close proximity
to enhance interedge backscattering.}
\label{schematic_illustration_0}
\end{figure}

{\it The model}---We consider here a QH system in which a point constriction with 
metallic gates, as in Fig.~\ref{schematic_illustration_0}, has been created.
The 2DEGs reside on the interface of two semiconductors,
the $xy$ plane ($z=0$) of 
Fig.~\ref{schematic_illustration_0}, and the gates over the top, a distance $10\,l_B$ ($l_B \approx 100$\,\AA, the magnetic length) from
the 2DEGs' plane, along the $y$ axis of Fig.~\ref{schematic_illustration_0}.
We assume here that the system is homogeneous 
along the gates, the $y$ direction, 
and that the gates and the 2DEGs are coplanar, 
an assumption justified at high $\left|V_\abbr{g}\right|$. 
As $\left|V_\abbr{g}\right|$ is increased, the electrons are repelled
leaving behind positively charged dielectric stripes of lateral extension
$a < |x| < b$, where $2a$ is the width of the gate region.
In contrast to edges created by cleaved edge barriers \cite{kang} of width
$\approx l_B$, the ones induced by gates are
much wider, on the order of $\approx 30\,l_B$,
with width growing quasi-linearly on $\left|V_\abbr{g}\right|$. 
At higher $\left|V_\abbr{g}\right|$ 
the edge structure becomes more complex than the simple single-channel chiral edge model. 
In absence of a magnetic field the configuration of the
electrostatic potential in
the 2DEGs' plane is $\phi(x)=V_\abbr{g}$ in the
gate region $|x|<a$ and $\phi(x)=V_1$ in the 2DEGs' regions, $|x|>b$. 
In addition, the positively charged dielectric in the depleted regions
creates an electric field $E^{\rm str.}_z$ proportional to
$\p_z \phi(x,z) =4\pi e n_0 /\epsilon=\tau$ (in \cite{pellegrini} donor charge density is
$n_0\approx 10^{11}\,{\rm cm^{-2}}$,
and $\epsilon=12.6$).
\looseness=-1

\begin{figure}
\begin{center}
\unitlength=1mm
\begin{picture}(86,30)
\put(0,16){\epsfig{file=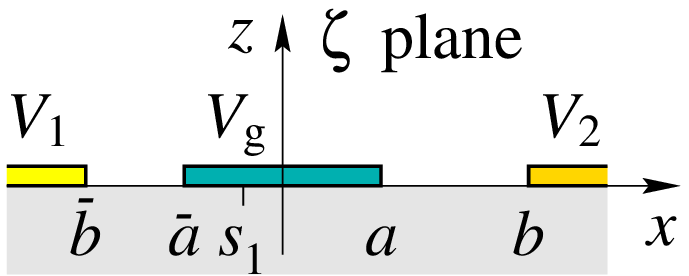,width=36.7mm}}
\put(0,0){\epsfig{file=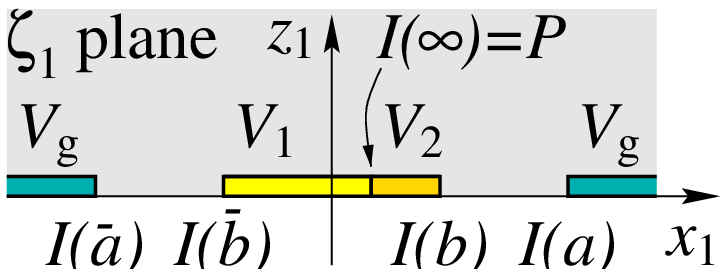,width=36mm}}
\put(38,-2){\epsfig{file=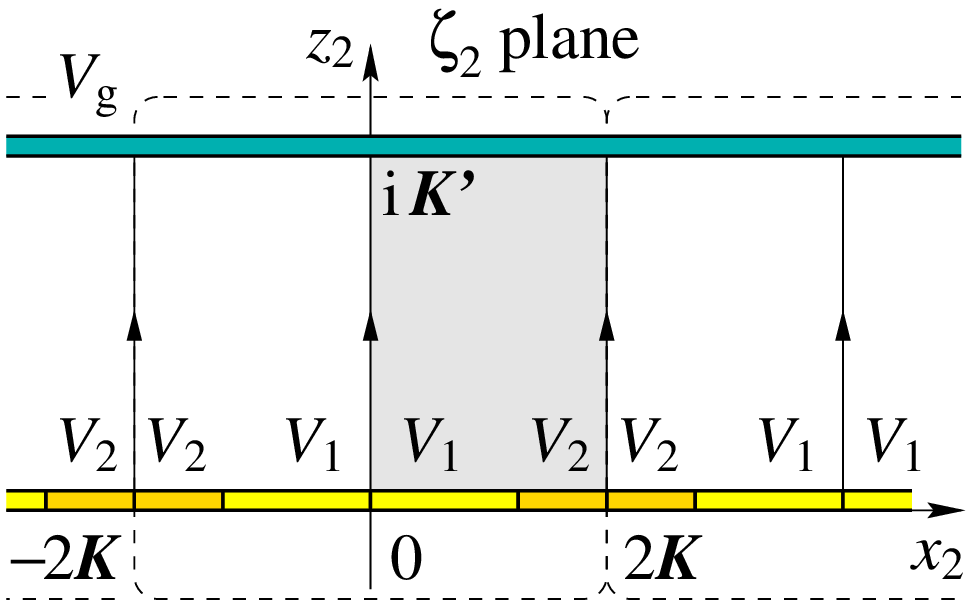,width=48mm}}
\end{picture}\nolinebreak
\end{center}
\vskip-\lastskip
\caption{
Upper left: original potential configuration in space -- metallic gate and 2DEG potential.
Since $V_1\neq V_2$ the distance 2DEGs--gate is different. Lower left: after
inversion around a point $s_1$ the 2DEGs' distances from the gate are equal. 
Right: forming a box configuration (shaded) by mapping the upper $\zeta_1$ half-plane, which in turn
maps the lower half $\zeta$ plane (\ref{z_to_z2}).
Application of the Schwarz mirroring principle about the vertical borders
of the shaded box leads to a periodic structure along the $x_2$ axis.
}
\label{inverted_system}
\end{figure}

{\it Nature of interactions}---Intuitively one expects  the presence of image
charges in the gate, halfway between edges, to  weaken the interedge Coulomb
repulsions or even to reverse their sign. The system examined here,
however, allows for exact calculations. To this end 
it is essential to allow for the 2DEG subsystems be at different potentials $V_1$ and $V_2$.
The 2DEG edges $\bar b,b$ therefore are placed at different distances from the gate. 
The gate extends in $\bar a < x < a$,
as illustrated in Fig.\ \ref{inverted_system} (upper left).
\looseness=-1

To calculate the field created by the metallic components we use
two conformal transformations. First, to get around the problem of
$\bar a-\bar b\ne b-a$, we use an inversion $I$ around a point $s_1$.
The latter is defined such that in the 
inverted space the 2DEGs are at same distance from the gate, 
see Fig.\ \ref{inverted_system} (left). 
Consecutively we use $\pi/2$ folding transformations at points
$I({\bar a}), I({\bar b}), I(b), I(a)$ as follows, 
\bea
\zeta_1 =\frac{R^2}{\zeta-s_1}+P \;\; , \;\;
\zeta_2 ={\rm sn}^{-1}\left(\frac{\zeta_1}{I(b)},\frac{I(b)}{I(a)}\right) + \bK
\;\; ,
\label{z_to_z2}
\eea
where $\zeta_1 = I(\zeta)$, $\zeta=x+\imu z$ etc., 
mapping the shaded parts of the spaces, see  Fig.\ \ref{inverted_system},
onto each other.
The radius of inversion, $R$, and the shift $P$ just 
play intermediary roles yielding a scale-invariant potential.
$\bK$, $\bK'$, $\bE$, $\bE'$ are the complete (complementary) Jacobi
elliptic integrals of $1^{\rm st}$ and $2^{\rm nd}$ kind and 
${\rm sn}^{-1}$ the inverse Jacobi elliptic function, each of modulus
$k=I(b)/I(a)$, see also below. 
\looseness=-1

To preserve the boundary conditions for the electric fields,
i.e., $E_z^{\rm el.}(x;z=0)=0$
in the depleted regions $a<x<b$, $\bar b < x < \bar a$ 
of the original problem,
we form the periodic configuration shown on
Fig.~\ref{inverted_system} (right), obtaining at the box sides
$E_{x_2}^{\rm el.}(x_2=\pm 2n\bK;z_2)=0$, $n$ integer.

The expression for the potential field
$\phi^{\rm el.}(\zeta_2,{\bar \zeta}_2)$ can be calculated analytically,
by summing up the series solution for the Laplace equation, 
to the close form
\begin{eqnarray}
 \label{close_form}
 \lefteqn{
 \phi^{\rm el.}
  = \tilde V_2-\left(V_\abbr{g}-\tilde V_2
  -\frac{\tilde V_1-\tilde V_2}{2\bK}C\right)
    \frac{\RE\{\imu \zeta_2\}}{\bK'}
 } \nonumber\\
 && {}+\frac{\tilde V_1-\tilde V_2}{\pi}
    \RE\left\{\imu\ln\frac{\vartheta_1\left(\frac{\pi}{4\bK}(\zeta_2+C),
    \eun^{-\frac{\pi\bK'}{2\bK}}\right)}
   {\vartheta_1\left(\frac{\pi}{4\bK}(\zeta_2-C),
   \eun^{-\frac{\pi\bK'}{2\bK}}\right)}
    \right\}
\; , \quad 
\end{eqnarray}
where $\vartheta_1$ is a Jacobi theta function of modulus
$k=S_k^-/S_k^+$ with
$S_k^\pm=\left[{(b-\bar a)(a-\bar b)}\right]^{1/2}
\allowbreak\pm\allowbreak \left[{(b-a)(\bar a-\bar b)}\right]^{1/2}$,
and $C=\zeta_2(I(\infty))$.
Here $\tilde{V}_i = V_i-A$, $i=1,2$, with the shift caused by $E_x^{\rm str.}$ \cite{def_A}. 
Due to
the vanishing component $E_{x_2}^{\rm el.}$
along the lines
$\zeta_2 = \pm 2n\bK+\imu z_2$, the Jacobian of the 
transformation $\zeta_2\to \zeta$ simplifies to  $J(2\to 0)= -S_k^+/[2D(\zeta)]$, and in the 
unfolded space
$E^{\rm el.}_x(x,z=0) = \p_{z_2} \phi^{\rm el.}(x_2,z_2) S_k^+/[2D(\zeta)]$,
with $D(\zeta)$ given in  \cite{D_F}.

For the calculation of the field created by the dielectric stripes
we generalize an idea given by Glazman and Larkin \cite{Glm+Lkn}
by representing the potential $\phi$ as the imaginary part of an
analytic function $F$, which, by using the boundary conditions for
its derivative, can be related to another function,
$f(\zeta)=\imu D(\zeta)(\rd F/\rd\zeta)$, 
whose imaginary part we would know everywhere along the $x$ axis.
By use of the Schwarz theorem one can generate 
$f(\zeta)$ on the whole plane.
Here,
for the uniformly charged asymmetric dielectric stripes one gets
$f(\zeta)=\tau[\imu D(\zeta) + {\cal F}(\zeta)]$, with $D(\zeta)$ and ${\cal F}(\zeta)$ given by
 \cite{D_F}.
 One finds $\phi$ by taking the opposite steps, leading immediately
 to the value of the $x$ component of the electric field,
 $E^{\rm str.}_x=\tau {\cal F}(x)/D(x)$.

From the equilibrium conditions of
$E_x^{\rm str.}(x,z=0)+E_x^{\rm el.}(x,z=0) \to 0$ 
at $x\to \bar b+0$ and  $x\to b - 0$ 
one obtains the edge-position equations 
\begin{figure}
\begin{center}
\epsfig{file=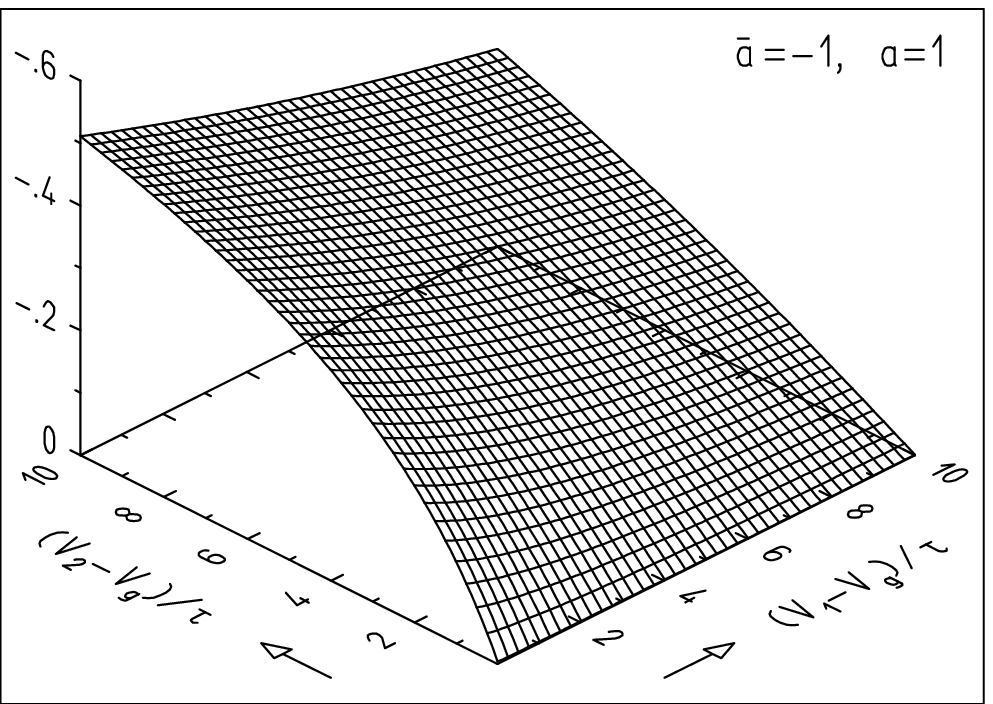,height=35mm}\hfill
\epsfig{file=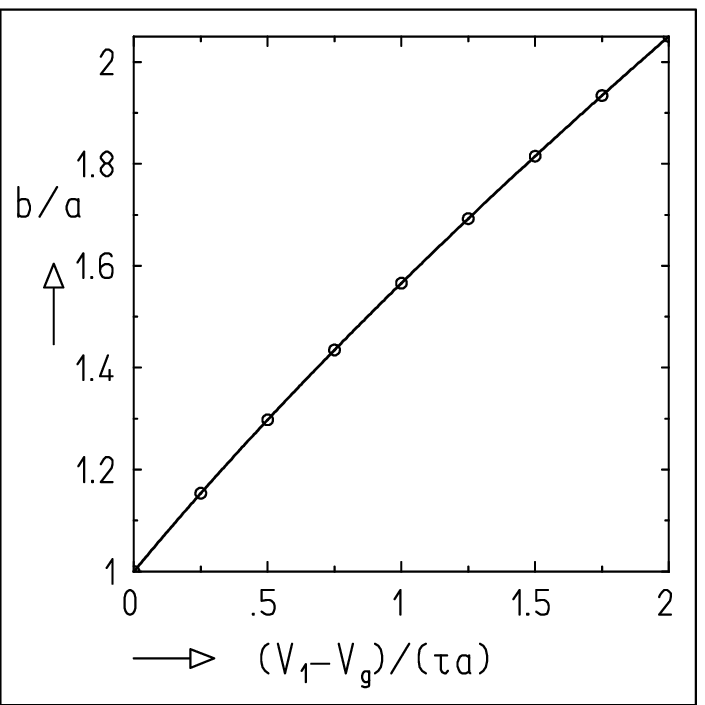,height=35mm}
\end{center}
\vskip-\lastskip
\caption{Left: $\partial_{V_1} b/\partial_{V_2} b $, 
the ratio of the differential
dependences of the edge position on the potentials on both sides
of the gate. Right: dependence of the edge position $b$ on
$V_1-V_\abbr{g}$ for the symmetric case ($V_2=V_1$; $\bar a=-a, \bar b=-b$).
}
\label{fig:d-pd-V1,V2}
\end{figure}
\bea
  \label{position_1}
  \!&\!\!&\frac{b+\bar b}{2}=\frac{a+\bar a}{2}-\frac{V_1-V_2}{\pi\tau} 
  \,, \,
  \\
  \label{position_2}
 \!&\!\!&  \frac{\tau}{2}(b-\bar b)^2  = 
  \frac{\tau}{2} \frac{\left({S_k^+}\right)^2\bE'}{\bK'}
  -\frac{S_k^+ V_0}{\bK'}-\frac{V_1-V_2}{\pi} l
\; , \qquad
\eea
where  $V_0 =V_\abbr{g} -V_2 -(V_1-V_2)C/(2\bK)$, and  
with $l$ \cite{def_l}.
Treating numerically the transcendental equation for $\bar b,b$
to obtain $\partial_{V_1} b/\partial_{V_2} b$,
we find that interactions remain repulsive at 
higher $|V_{\abbr g}|$ as they do for low $|V_{\abbr g}|$, Fig.~\ref{fig:d-pd-V1,V2} (left). 

The edge position $b$, when $V_2=V_1$, 
increases quasi-linearly with  $|V_\abbr{g}|$, 
$ b \approx {\kappa a}[{\ln \left(4\kappa/{\rm e}\right)
-\ln\ln \left(4\kappa/{\rm e}\right)}]^{-1}\,$,  
where $\kappa = (V_1-V_\abbr{g})/\tau a$, 
as shown on Fig.~\ref{fig:d-pd-V1,V2} (right).

{\it Edge structure}---In the edge of a bulk integer filling $\nu=1$ 
there exists a hierarchy of FQH energy gaps
at fillings $f=p/(2mp\pm 1)$, for integers $m$ and $p$. 
The system lowers its energy \cite{glazman,chklovskii,beenaker,chang}
by relocating some electrons from higher to
lower energies forming fractional-filling ISs
around the points $x_{f\pm} = \pm b [(1-k^2f^2)/(1-f^2)]^{1/2}$ of filling 
$n(x_{f\pm}) = f n_0$ where $n(x) = n_0[(x^2-b^2)/(x^2-a^2)]^{1/2}$.
While a large number of alternating ISs and CSs can in principle form at the edge,
only the widest few will be realized in practice.
The largest charge gaps are known to occur at $f=1/3$, $2/3$ \cite{chang} 
and we calculate here the widths $\Delta a_{1/3}$, $\Delta a_{2/3}$.
\looseness=-1

In presence of a magnetic field,
potential differences of value \cite{chang,glazman}
$V_f=1.02\,$mV$ \ll \left|V_1-V_\abbr{g}\right|$ 
on the sides of the ISs $f=1/3$, $2/3$ arise. 
The charge density in the region of ISs, ${\cal N}(x) = n(x) - n_f$, 
should now be treated as $x$-dependent. 
Since $V_f/\left|V_1-V_\abbr{g}\right|\approx 10^{-3}$ \cite{pellegrini}, 
we can expand
the even function $n(x)$ around the points $x_{f\pm}$ in one even polynomial
$\sum_{i=0}^N \tilde n_i^{(N)} x^{2i}$ matching $n(x)$ up to the
$N$th derivative there. For $N=2$, e.g., one gets
$\{\tilde n_i^{(2)}\} = \{n_f-\frac{5}{8}x_f n_f'+\frac{1}{8}x_f^2n_f'',
\frac{3}{4}x_f^{-1} n_f'-\frac{1}{4}n_f'',
-\frac{1}{8}x_f^{-3} n_f'+\frac{1}{8}x_f^{-2} n_f''\}$.

\begin{figure}
\unitlength=1mm
\begin{center}
\epsfig{file=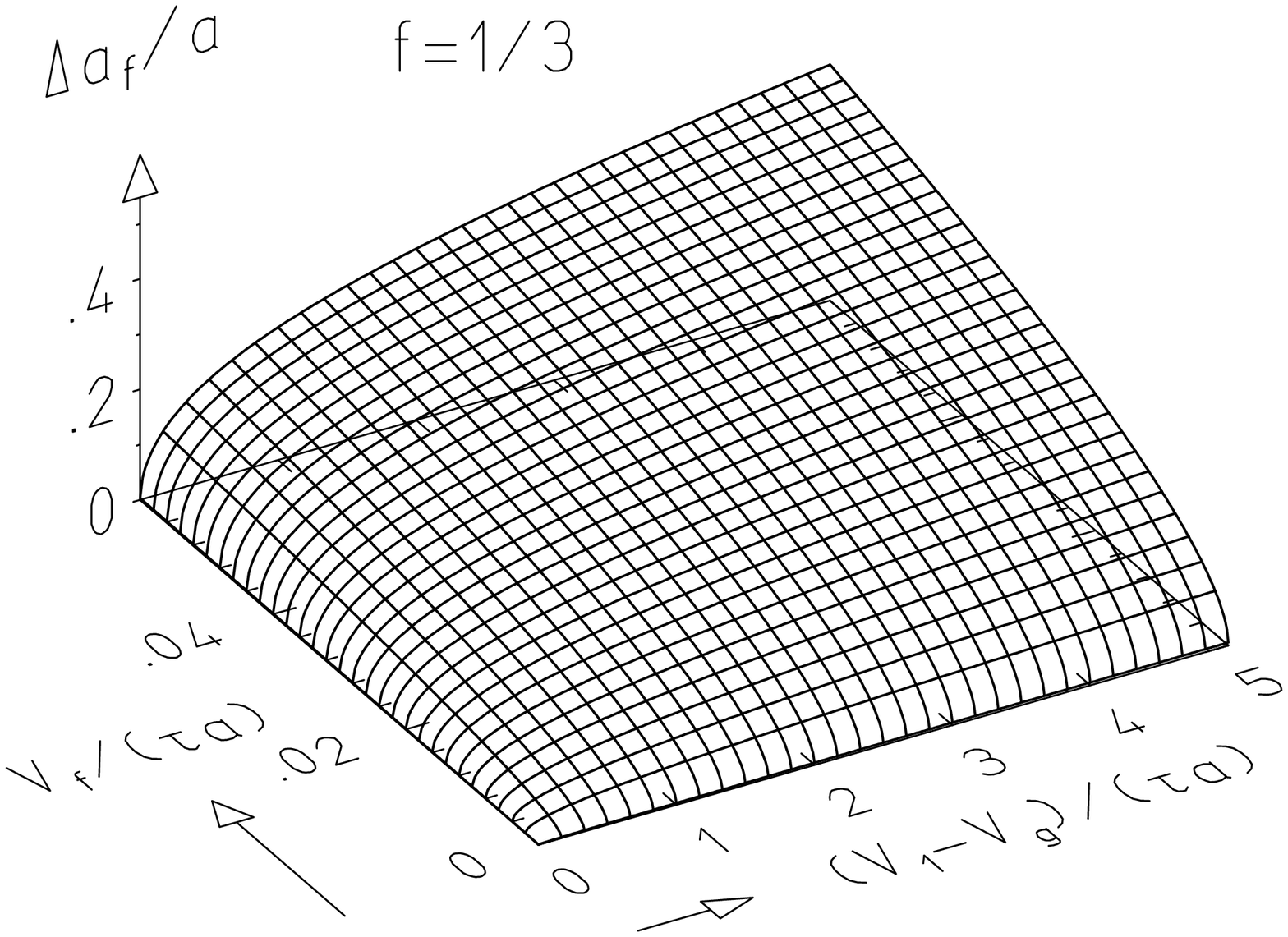,width=42mm}
\epsfig{file=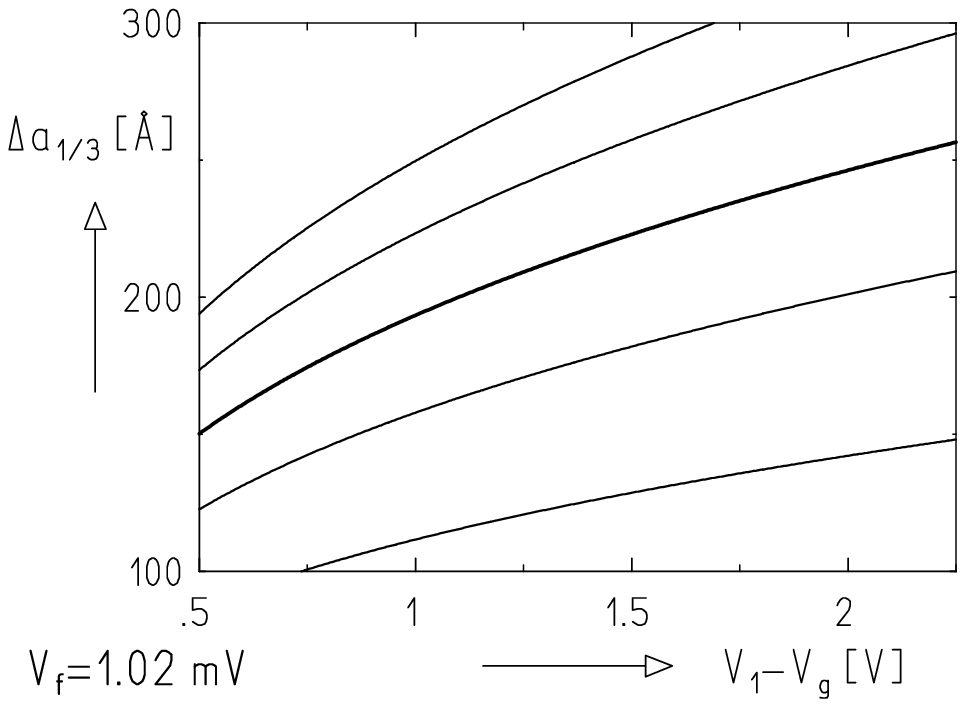,height=30mm}\hfill
\end{center}
\vskip-\lastskip
\caption{
Left: IS width $\Delta a_{1/3}$, in units of half-gate width, $a$.
Right: IS width $\Delta a_{1/3}$, in \AA\ as function of
$V_1-V_\abbr{g}$ and $V_f$ for the first-order approximation
and the values $V_f\to (i/3)V_f$, $i=1,\ldots,5$, from bottom to top 
(in \cite{pellegrini}, $\left|V_1-V_\abbr{g}\right|=0.7\,$V, $a=3000$\,\AA).
}
\label{stripe_widths}
\end{figure}

\begin{figure}
\begin{center}
\unitlength=1mm
\begin{picture}(86,34.7) 
\put(0,0){\epsfig{file=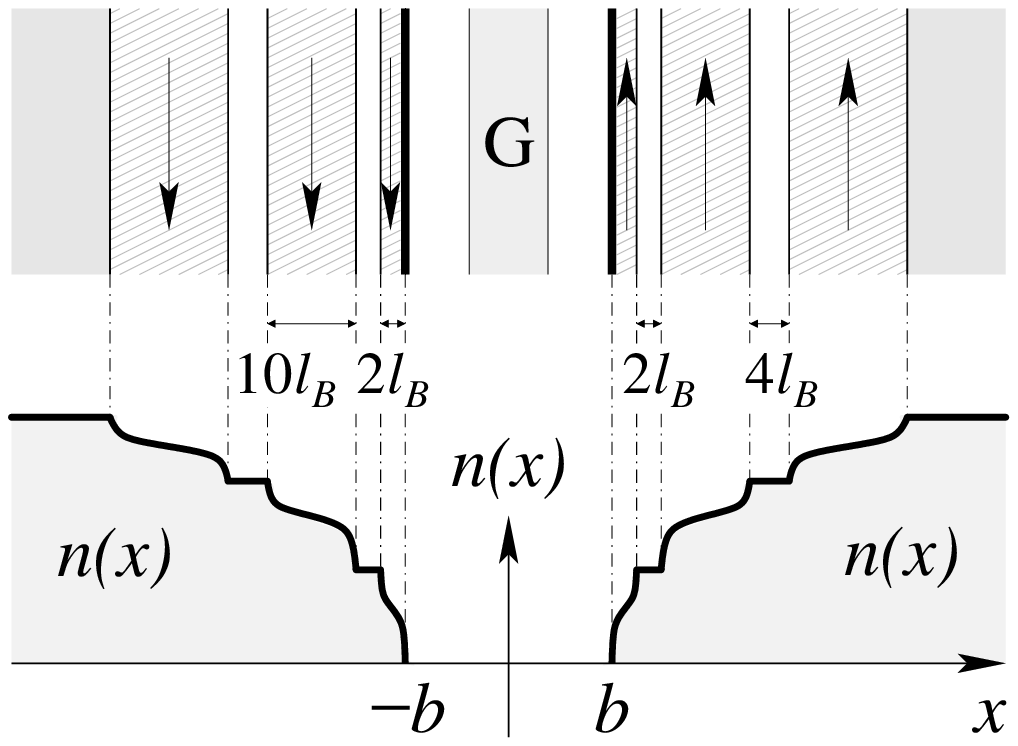,width=47mm}}
\put(50,0){\epsfig{file=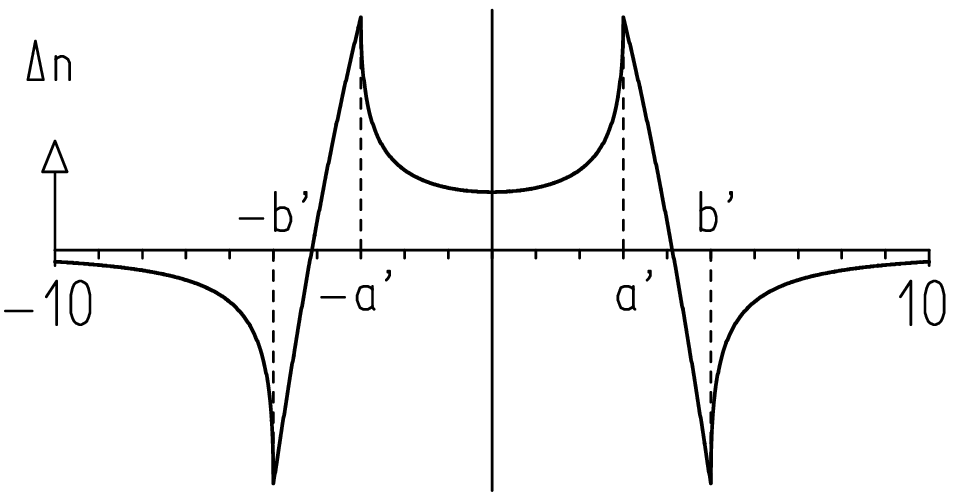,width=36mm}}
\end{picture}\nolinebreak
\end{center}
\vskip-\lastskip
\caption{Left: structure of the edge of a QH system surrounding a
split gate, in presence of a strong magnetic field. 
Shaded areas notify current-carrying CSs, separated by insulating ISs. 
Right: magnified plot of additional electron charge caused by the 
magnetic field around the points $x_{f\pm}$ on the edge profile, 
$\Delta n = (n_f'/2x_f)[(a'^2+b'^2)/2-x^2\pm |D'(x)|]$, $(+)$ for
$|x|>b'$, $(-)$ for $|x|<a'$, and zero otherwise. 
}
\label{charge_striped_magn}
\end{figure}

The potential field created by the charged stripes extended along
$a'<|x|<b'$ with $a'>b$, found by integrating $f(\zeta)/D'(\zeta)$, 
$D'(\zeta) := D(\zeta)\{\bar b,\bar a,a,b\to -b',-a',\allowbreak a',b'\}$, 
along the real axis
(of the general form $f(\zeta) = \tau [\imu D'(\zeta)
+ {\cal F}(\zeta,a',b')]$), is now modified to account in principle
for the $n$th order of approximation for ${\cal N}(x)$
[or $\tau(x) = (2\pi e/\epsilon) {\cal N}(x)$],
$\tau_n(x) = \tau_n x^n$, giving for ${\cal F}$ 
\bea
  {\cal F}_n = 
   \tau_n\zeta^{n+2}
\sum_{k=0}^{\left[n/2+1\right]}
   C_k^{-\frac{1}{2}}\left({\textstyle \frac{a'^2+b'^2}{2a'b'}}\right)
   \left(\frac{a'b'}{\zeta^2}\right)^k \quad ,
\eea
where $C_k^{-\frac{1}{2}}(t)$ is a Gegenbauer polynomial, 
and $[\cdot]$ represents the integer part
of a real number. 

To second order, from conditions of  equilibrium at both edges of the stripes, we 
find that $a'$, $b'$, 
in addition to the condition $a'^2+b'^2=2x_f^2$, fulfill also
\bea
  \frac{2-\bar k'^2}{2}\bE'
  -(1-\bar k'^2)\bK'=\frac{3\sqrt{2}}{4}(2-\bar k'^2)^{3/2}
  \frac{V_f}{\tau'x_f^2} \quad ,
\eea
where $\bar k=a'/b'$, $\bar k'^2= 1 - {\bar k}^2 $.
For the IS 
widths $\Delta a =b'-a'$ we find
$\Delta a_{f} \approx \left[(16/\pi \tau){f V_f}/{(1-f^2)^{3/2}} \right]^{{1}/{2}} b^{{1}/{2}}$,
proportional to $V_\abbr{g}^{1/2}$ up to logarithmic corrections.
Numerical results for $\Delta a_f$ are represented in Fig.~\ref{stripe_widths}. 
The narrower outer ISs 
are of size 180--200\,\AA, 
corrected upward in higher
orders of expansion on $n(x)$. 
The wider inner $2/3$ channels, of size $\approx 400$\,\AA, 
are  $\approx 1000$\,\AA\  away. 
These ($2\,l_B$, $4\,l_B$) are 
distances for which the overlap of electron wave functions on opposite sides of ISs is exponentially
small. Therefore even the $1/3$ IS is 
insulating in the transversal direction at high $\left|V_{\abbr g}\right|$.
\looseness=-1

{\it Stability of edge structure}---The dynamics of wide edges in strong magnetic fields, when 
the emerging ISs were assumed to be narrow, 
was studied in \cite{vignale2}.
The excitation spectrum was found to be composed of 
 an edge magnetoplasmon and slower-moving acoustic modes. In the latter 
 charge density oscillates across the profile of the edge ($x$ axis).
 The geometry of our system and the experimental values of $V_{\abbr g}$ however are such that
 the 1/3 and 2/3 ISs of widths $2\,l_B$, $4\,l_B$ are of comparable size to the 
respective 1/3 and 2/3 CSs ($2\,l_B$, $10\,l_B$ wide), Fig.~\ref{charge_striped_magn},
 and the model of coupled chiral LLs, is adopted below.

\begin{figure}
\unitlength=1mm
\begin{center}
\epsfig{file=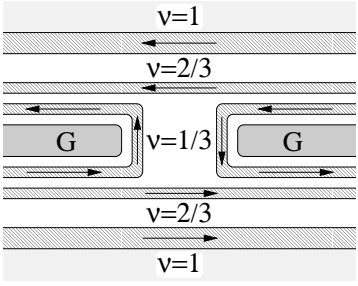,clip,width=50mm} 
\end{center}
\vskip-\lastskip
\caption{
Schematic illustration of high $\left|V_\abbr{g}\right|$ splitting
of an $\nu=1$ edge into channels of fractional filling in QH
systems with constriction, similar to the experiments of Ref.~\cite{pellegrini}.
}
\label{edge_config}
\end{figure}

First of all, since the ISs, illustrated in Fig.~\ref{edge_config}, 
are wide, the effective statistics of fluctuating channels
is the one characterizing the $\nu=1/3$ edges \cite{vignale}.
In the absence of inter-channel Coulomb interactions, the
LL formed between $\nu=1/3$ channels at the line
junction anticrossing the Hall bar is equivalent to a 
LL along the junction at filling $1/\nu$ \cite{renn,papa},
a reflection of quasiparticle--electron duality in tunneling processes.  
Starting with an open constriction, slowly turning off $V$, the system flows from relevant 
quasiparticle backscattering, suppressing the 
source-to-drain conductance according to
$G- \nu e^2/h \sim - V^{2\nu-2}$, 
to a closed constriction with irrelevant electron tunneling with $G\sim V^{2/\nu-2}$, 
unable to open the constriction at small $V$.
The system is insulating at low bias 
with $I$--$V$ characteristics of a QH system of bulk $\nu=1/3$,
in  agreement with observations of Ref.~\cite{pellegrini}.

Accounting for the inter-channel Coulomb interactions
the action of same-chirality branches is
\bea
 S = \sum_{i,j=1}^3 \int  \rd x \, \rd \tau 
  \left[ \frac{2\imu\delta_{ij}}{\nu}
  \,\p_\tau \phi_i \, \p_x \phi_j + V_{ij} \,
  \p_x \phi_i \, \p_x \phi_j \right]
  \; , \;\,
\eea
where  $\phi_i$, $i=1,2,3$, are bosonic fields associated with each CS. 
$\p_x \phi_i =-\sqrt{\pi}\rho_i(x)$ is the charge density of channel $i$ and quantization is assumed
based on the commutation relations
$[\rho_i(x),\rho_j(x')]=-\frac{\imu\nu}{2\pi}\delta_{ij}
\p_x\delta(x-x')$.
We can assume in general the intra-channel interactions $V_{ii}$
for $i=1,2,3$ to 
be different, and the inter-channel ones to fulfill $V_{ij}=V_{ji} \neq V_{ii}$.
The diagonalized action can be written  as 
\looseness=-1
\bea
S= \sum_{i=1}^3 \int \rd x \, \rd \tau
\left[\frac{2\imu}{\nu} (\p_\tau {\varphi}_i)(\p_x {\varphi}_i)
 + \lambda_{i} (\p_x {\varphi}_i)^2  \right]  \; , 
\eea
where ${\varphi}_i=M_{ij}\phi_j$ and $\lambda_i$ are eigenvalues of $V$. 
Since all channels are of same chirality,
the interaction-dependent $M$ matrix is 
unitary, $(M^\dagger M)_{ij}=\delta_{ij}$,
as opposed to $M^\dagger J  M = J$, with $(J)_{ij}=\delta_{ij}{\rm sign}(j)$
in presence of opposite-chirality channels,
resulting in $(M^\dagger M)_{ij} \neq \delta_{ij}$.
The correlation functions between electrons, 
$R_i^{\rm el.} \sim \allowbreak \eun^{\imu \sqrt{4\pi}\phi_i/\nu}$,
of branch $i$,  
have the form $\bigl< {R_i^{\rm el.}}^\dagger (\tau) R_i^{\rm el.}(\tau')\bigr>
\sim \allowbreak \prod_{j=1}^{3}(\lambda_i (\tau -\tau'))^{-2 (M^\dagger M)_{ij}/\nu}$, 
and the scaling dimensions are $d_i=\sum_j (M^\dagger M)_{ij}/\nu=1/\nu$, irrelevant for $\nu=1/3$
and unaffected by the interactions.

Another point of concern is
the question of stability of the channel structure under inter-channel tunneling processes
mediated by impurities along the edge. 
For repulsions $V_{ij}=V$  ($\le V_{ii}=U$),  
the action is diagonalized in terms of two neutral and one charged field.
At low energies the neutral fields tend to antilock charge densities in the consecutive
channels. 
Inter-channel tunneling processes lead to the chiral sine-Gordon model
$\cos(\frac{\sqrt{8\pi}}{\nu}\varphi_n)$, $n=1,2,$ for neutral fields.
The ${\it cos}$ operator is relevant only for $\beta^2 \le 16\pi$ \cite{sondhi}, 
a condition not realized in these systems since $\beta^2=8\pi/\nu^2$
and $\nu =1/3$. 

As $\left|V_\abbr{g}\right|$ is lowered we expect the $1/3$ IS channel to eventually close
and the remaining $2/3$ IS to extend under the gate. 
At large bias, the same argument as in {\it high\/} $\left|V_\abbr{g}\right|$ for the conductance applies,
with $G - \nu e^2/h \sim - V^{2\nu -2}$ and $\nu=2/3$.
At low $V$, however, Coulomb interactions between oppo\-site-chirality channels as discussed
in \cite{papa} are expected to be strong leading to perfect transmission through 
the constriction and exhibiting the symmetry $\nu=1/3 \leftrightarrow \nu=2/3$ between 
the low and high $|V_{\abbr g}|$ observed in the experiments of Ref. \cite{pellegrini}. 

We are grateful to M. Grayson,
V. Pellegrini, M. Polini, S. Roddaro for discussions and to  
A. MacDonald and A. Tsvelik for comments on the manuscript 
and discussions. 
This work was supported by NSF grant DMR-0412956.


\begin{thebibliography}{50}
\bibitem{D_F} The functions $D$ and ${\cal F}$ for
asymmetric dielectric stripes are
$
D(\zeta)=[(\zeta-\bar b)(\zeta-\bar a)(\zeta-a)(b-\zeta)]^{1/2}$ 
and 
$
{\cal F}(\zeta) =
\thalf(\zeta-\bar b)(\zeta-b)+\thalf(\zeta-\bar a)(\zeta-a)
-{\textstyle\frac{1}{8}}(b-a-\bar a+\bar b)^2$.
This ${\cal F}$ is valid only for uniformly charged stripes.
\bibitem{def_l}
The length $l$ scaling the potential difference in (\ref{position_2}) is
$  l=b-\bar b-2 \bigl[(a-\bar b)(\bar a-\bar b)\bigr]^{1/2} 
  -S_k^+\left(2E\left(\am\left(\frac{C}{2}\right)\right)
  -\frac{C\bE}{\bK}\right) \,.
$
\end{thebibliography}
\end{document}